\documentclass[a4paper,11pt]{article}
\pdfoutput=1 

\usepackage{jinstpub} 

\usepackage{lineno}
\usepackage{listings}
\usepackage{xcolor}

\title{\boldmath RooHammerModel: interfacing the HAMMER software tool with HistFactory and RooFit}

\author[a,b,1]{J. Garc\'ia Pardi\~nas,\note{Contact author.}}
\author[a,b,2]{S. Meloni,\note{Contact author.}}
\author[c,3]{L. Grillo, \note{Now at School of Physics and Astronomy, University of Glasgow, Glasgow, United Kingdom.}}
\author[d]{P. Owen}
\author[a,b]{M. Calvi}
\author[d]{N. Serra}

\affiliation[a]{Universit\`a di Milano Bicocca,\\Milano, Italy}
\affiliation[b]{INFN Sezione di Milano-Bicocca,\\Milano, Italy}
\affiliation[c]{School of Physics and Astronomy, University of Manchester,\\Manchester, United Kingdom}
\affiliation[d]{Physik-Institut, Universit\"at Z\"urich,\\Z\"urich, Switzerland}

\emailAdd{julian.garcia.pardinas@cern.ch}
\emailAdd{simone.meloni@cern.ch}

\abstract{Recent $B$-physics results have sparkled great interest in the search for beyond-the-Standard-Model (BSM) physics in $b\to c\ell \bar{\nu}$ transitions.
The need to analyse in a consistent manner big datasets for these searches, using high-statistics Monte-Carlo (MC) samples, led to the development of HAMMER, a software tool which enables to perform a fast morphing of MC-derived templates to include BSM effects and/or alternative parameterisations of long-distance effects, avoiding the need to re-generate simulated samples. This paper describes the development of RooHammerModel, an interface between this tool and the commonly-used data-fitting frameworks HistFactory and RooFit. In this document, the structure and functionality of the user interface are explained. Information of a public repository where it can be accessed is provided, as well as validation and performance studies of the interface. The methods developed in the construction of RooHammerModel can provide useful information for alternative future attempts to interface HAMMER with other data-fitting frameworks.}

\keywords{Analysis and statistical methods, Data processing methods}

\arxivnumber{2007.12605} 

\notoc

\begin{document}
\maketitle
\flushbottom

\section{Introduction}
\label{sec:introduction}

The recent $B$-physics anomalies found in $b\to c \ell \bar{\nu}$ transitions~\cite{Lees:2012xj,Lees:2013uzd,Huschle:2015rga,Sato:2016svk,Hirose:2016wfn,Aaij:2015yra,Aaij:2017uff,Abdesselam:2019dgh} have generated great interest in the study of beyond-the-Standard-Model (BSM) effects in charged-current semileptonic decays, and both the LHCb and Belle II experiments have planned intensive physics programs~\cite{Bediaga:2018lhg,Kou:2018nap} to investigate the nature of these anomalies. The relevant decay modes are characterised by large branching ratios and consequently large signal yields. Precise measurements of these large datasets, require the production of enormous Monte Carlo (MC) samples.
As an example, for an ongoing LHCb analysis with about six million data events selected, a production of roughly three billion simulated events is needed. In order to perform self-consistent and up-to-date measurements, the simulation models should include an adequate description of BSM effects and a state-of-the-art parameterisation of soft QCD effects.
However, the high resource consumption of the simulation prevents the generation of devoted samples for every possible configuration of decay parameters.

The recently-developed HAMMER~\cite{Bernlochner:2020tfi} software tool allows to perform a fast re-weighting of a given MC sample to effectively change the decay matrix element into a new desired one, without the need to regenerate the MC events. The transformation can be done for any possible values of the parameters governing the decay amplitude, which opens the possibility of using HAMMER to measure those parameters in fits to collision data.
To enable the necessary dynamic re-weight of the MC distributions, an interface must be implemented between HAMMER and the used fitting framework.

Semileptonic charged-current decays are typically studied via binned template fits, and one of the most extended and user friendly frameworks that allows to do so in High-Energy Physics (HEP) is HistFactory~\cite{Cranmer:1456844}. This tool has a limited functionality for template morphing, only being able to provide interpolations between fixed-shape histograms that have to be provided as input. In order to use a generic shape parameterisation in HistFactory, in particular the functional forms provided by HAMMER, new structures have to be developed.

This paper presents the construction of RooHammerModel, a \texttt{C++} based interface between the HAMMER tool and the HistFactory and RooFit~\cite{verkerke2003roofit} fitting frameworks. The interface has been constructed to be fully flexible regarding the decay mode and parameter configuration, allowing the user to specify any of the possible configurations available in HAMMER. The technical details of the interaction with HAMMER are fully kept inside the class, providing a user experience which is as close as possible to that of a standard RooFit model for a probability density function. The interface has been optimised to minimise the fitting time and memory consumption and has been tested on a typical use case.

The structure of the paper is as follows. Section~\ref{sec:structure} describes the overall structure of the interface. Section~\ref{sec:user} explains the user configuration. Section~\ref{sec:technical} explains several aspects of the particular implementation and the designed solutions to some challenges that arose in the process. Section~\ref{sec:tests} presents the studies done to test the fitting interface. Section~\ref{sec:performance} shows a performance analysis. Finally, some concluding remarks are given in section~\ref{sec:conclusions}.
\section{Structure of the framework}
\label{sec:structure}

The usage of the interface assumes that the desired simulation samples have been previously pre-processed with the HAMMER tool (as explained in detail in Ref.~\cite{Bernlochner:2020tfi}) and that the corresponding results, collected in binary files, are available. The overall structure of the interface is similar to a standard HistFactory fitting script, but incorporating a new class, RooHammerModel, to model the desired decay components using HAMMER histograms. The technical interaction with HAMMER is implemented internally within the class, such that the user only needs to specify the desired model configuration and parameters. The general features of both the new class and the fitter structure are discussed in the following.

The RooHammerModel class inherits from \texttt{RooAbsPdf} and implements a normalised piecewise function of arbitrary shape, which is equivalent to a binned probability density function, PDF. It can be used both in HistFactory and in plain RooFit. The PDF represented by the class is currently defined in a three-dimensional variable space, although this functionality can be easily extended in the future to perform fits in a different number of dimensions. Since the shapes are obtained from the HAMMER tool, any arbitrary set of three variables in the decay can be adequately modelled by the class. The class presents some additional methods that allow the extension of typical HistFactory features to variable-shape histograms, as explained in section~\ref{sec:technical}.

The typical structure of a HistFactory fitting script can be maintained, with some additions needed to implement the full functionality of the RooHammerModel class.
\begin{itemize}
    \item As a first step, the set of templates needed to describe the physics model under consideration are loaded, as well as the data histograms to be fitted. The desired instances of the RooHammerModel class are created, and they are used to obtain associated initial templates, which are passed to the HistFactory model (as explained in section~\ref{sec:technical}).
    \item Afterwards, the HistFactory model is internally transformed into a RooFit probability density function (PDF), as usually done in this framework.
    \item The next key point is the substitution in the model of the relevant initial templates by the corresponding RooHammerModel objects, which enables the dynamical variation of the templates' shape.
    \item Finally, the fit to data is done with the usual commands.
\end{itemize}
\section{User interface}
\label{sec:user}

The decay-rate amplitudes contained in HAMMER make use of the Effective Field Theory, that encodes all the short-distance effects (including the BSM contributions) in the so called Wilson coefficients, and the soft-QCD effects in kinematic-dependent form factors~\cite{Bernlochner:2021vlv}. The functional form of the form factors depends on the chosen phenomenological parameterisation. 

The main idea behind HAMMER is that the differential decay rate can be expressed as a linear combination of terms that depend on the Wilson coefficients. Consequently, the 
transformation between distribution shapes that corresponds to a change in Wilson coefficients' values can be obtained via linear algebra operations. Furthermore, some of the form factor parameterisation classes implemented in HAMMER provide the functionality to reweight the distributions changing the value of the form factor parameters inside the model. These classes are denoted with ``Var'' in their name. In these classes, the form factors are linearised in the parameters' space around the chosen initial values. With this approximation, the decay rate for the process, as a function of any observable, takes a linear form also in the form factor parameters. More complete information can be found in~\cite{Bernlochner:2020tfi}.

In the interface described in this document, the handling of Wilson coefficients and form-factor parameters is done by the RooHammerModel class.

\subsection{RooHammerModel class}
\label{sec:user_RooHammerModel}

The user can include in the model as many HAMMER-derived decay templates as desired, by creating one instance of the RooHammerModel class for each of them. The class is configured for a particular decay via its constructor, shown below. The user has to provide the information of the decay process and the chosen form-factor parameterisation, matching the information that should have been passed to HAMMER at the MC pre-processing step. The user can also specify which of the Wilson coefficients and/or form-factor parameters are to be varied during fitting, by passing lists of arguments to the constructor.

\paragraph{Constructor of the class.}

The default constructor of RooHammerModel is shown below, together with a description of the arguments to be provided.

\begin{lstlisting}[language=C++]
  RooHammerModel(const char *name, const char *title,
              std::string_view WCprocessname,
              std::vector<std::string>* _WCparamnames,
              const RooArgList& _reWCparamlist,
              const RooArgList& _imWCparamlist,
              std::string_view FFprocessname,
              std::string_view FFmodelname,
              std::vector<std::string>* _FFparamnames,
              const RooArgList& _FFparamlist,
              std::vector<std::string>* filenames,
              std::string_view histoname_noerrors,
              std::string_view histoname_witherrors,
              std::string_view schemename);
\end{lstlisting}

\begin{itemize}
    \item \texttt{name} and \texttt{title} denote the usual name and description attributes of a \texttt{RooAbsPdf}.
    \item \texttt{WCprocessname} denotes the quark-level weak-decay process, following the HAMMER naming convention. For example, \textcolor{brown}{``BtoCTauNu''}.
    \item \texttt{\_WCparamnames} is a vector containing the names of the (subset of) Wilson coefficients to be floated in the fit, without any particular ordering. The naming convention is the same as in HAMMER.
    \item \texttt{\_reWCparamlist} and \texttt{\_imWCparamlist} contain, respectively, the lists of real and imaginary parts of the Wilson Coefficients whose names are given in \texttt{\_WCparamnames} (the ordering of the three lists has to match). The parameters can be either \texttt{RooRealVar} or \texttt{RooFormulaVar}.\footnote{Note that the \textit{name} and the \textit{value} fields, used to instantiate these RooFit variables, are not in any way related to the internal interaction with HAMMER and its naming conventions. This choice gives full freedom to the user in naming the observables.}
    \item \texttt{FFprocessname} denotes the hadronic decay process, following the HAMMER naming convention. For example \textcolor{brown}{``BtoD*''}.
    \item \texttt{FFmodelname} denotes the chosen form-factor parameterisation, using the HAMMER naming convention. For example, \textcolor{brown}{``BGL''}.
    \item \texttt{\_FFparamnames} is a vector containing the names of the (subset of) form-factor parameters to be floated in the fit, without any particular ordering. The naming convention is the same as in HAMMER.
    \item \texttt{\_FFparamlist} contains the list of form-factor parameters matching the names given in \texttt{\_FFparamnames}. The parameters can be either \texttt{RooRealVar} or \texttt{RooFormulaVar}.
    \item \texttt{filenames} is a vector of names for the HAMMER buffer files to be used. If the vector contains more than one element, the files will be combined inside the class. This behaviour is aimed at combining the sub-files obtained from a separate pre-processing of different sub-samples of the MC dataset for a given specie.
    \item \texttt{histoname\_noerrors} denotes a user-defined name, chosen at HAMMER pre-processing time, given to the HAMMER histogram object that constitutes the core of the PDF. This histogram does not have errors associated to the bins, only central values. The motivation for using two versions of the histogram, with and without bin errors, is explained in section~\ref{sec:technical}.
    \item \texttt{histoname\_witherrors} denotes a user-defined name, chosen at HAMMER pre-processing time, given to a HAMMER histogram object that contains both bin errors and central values. The usage of this histogram is explained in section~\ref{sec:technical}.
    \item \texttt{schemename} denotes a user name given to the form-factor scheme, chosen at HAMMER pre-processing time.
\end{itemize}

Note that the constructor above is generic to any decay channel and decay-amplitude parameterisation. The possibility to pass \texttt{RooFormulaVars} as observables to the model is particularly interesting in an analysis. A specific use-case scenario, for example, could be the need to include some relationship between Wilson coefficients to perform a model-dependent fit.

\subsection{Fitting script}
\label{Sec:Fitting_script}
The RooHammerModel interface can be employed as a standalone RooFit object, and is therefore usable in any program developed with the RooFit package. However, it has been developed with particular focus on the HistFactory package and an example script is provided as a guideline to perform a standard fit.

In this script, a model with two decay channels is configured. The shape of one of the two decay channels can vary according to the \textit{CLNVar} HAMMER form factor parameterisation. The construction of the part of the fitting model that does not depend on HAMMER and the overall manipulation of the likelihood can be done as usual in HistFactory.

A particularly interesting HistFactory feature that has been extended to the new scheme is the treatment of model-template uncertainties, using the Barlow-Beeston Lite method~\cite{Barlow:1993dm}. Both the fitting script and the RooHammerModel have been adapted to be able to use the bin errors computed by HAMMER, as explained in section~\ref{sec:technical}. The inclusion of template-uncertainties in the model can be activated as usual, via \texttt{RooStats::HistFactory::Sample::ActivateStatError}.

\section{Technical implementation}
\label{sec:technical}

This section describes the strategy and techniques adopted to overcome several challenges identified during the process. These methods can also be used to develop alternative interfaces of HAMMER with other fitting frameworks.

\subsection{Handling of the HAMMER objects}

To reduce the memory consumption, the number of \texttt{HAMMER::HAMMER} instances loaded in memory while fitting is kept to only one for each specie in the model. That single instance is created at construction time, and kept as a data member of the associated RooHammerModel. Any copy of a particular RooHammerModel object attempted in the fitting framework will not result in the creation of a new \texttt{HAMMER::HAMMER} instance, but on the creation of a pointer to the instance stored in the original RooHammerModel object. The internal usage of thread-safe standard-library pointers, ensures that our class can be safely used in a multi-threaded environment.

\subsection{Histogram caching}

To avoid unnecessary computations during fitting and speed-up the process, the three-dimensional histogram that represents the model obtained from HAMMER for a given decay channel is only updated if any of the Wilson coefficients or form-factor parameter values is changed at the current minimisation step. Otherwise, no reweight is performed and a cached version of the previous histogram is used.

\subsection{Treatment of empty bins}

If the number of events in the MC sample used to model a component in HAMMER is low for a given binning scheme, it may happen that some bins are empty, even if the underlying model for those bins is not zero. This would be problematic in a fit using HistFactory if the data sample to be fitted happens to have some events in the affected bins. To prevent this potential situation, every histogram retrieved from HAMMER is automatically corrected for empty bins inside RooHammerModel. This correction is implemented by adding to the nominal histogram a constant flat histogram where the bin content value is an arbitrarily small number
\cite{Aaij:2015yra}, conventionally set to $10^{-10}$. Note that this approach avoids an iteration over the bins to check whether or not they are empty, and the effect on the non-empty bins is negligible. This correction is applied whenever the model is reweighted.

\subsection{Model normalisation}

To preserve the statistical definition of PDF, the histogram that represents the physical distribution in RooHammerModel is always re-normalised after any change in shape, such that its integral on the defined phase space is always equal to one. The re-normalisation is done after correcting for empty bins. Note that this normalisation makes the RooHammerModel distribution insensitive to any change in a HAMMER parameter that represents an overall multiplicative factor to the total amplitude. If such a parameter exists in the desired model, the associated object passed to RooHammerModel must be constant.\footnote{If relevant, the parameter representing an overall multiplicative factor can be dealt with externally in the fitting framework, forming part of a yield parameter that multiplies the RooHammerModel PDF.}. Apart from the automatic model normalisation, the RooHammerModel class provides all needed analytical integrals in any region of the phase space and for any desired dimensionality up to 3D.

\subsection{Template uncertainties in the fit}
\label{sec:technical_BB}

The limited size of the MC samples used to describe the model in the fit intruduces a source of systematic uncertainty on the fit parameters. One way to include this uncertainty automatically in the fit is the Barlow-Beeston method~\cite{Barlow:1993dm}. In this approach, the logarithmic likelihood used for the fit is supplemented with constraints that depend on values and statistical errors for each bin in each of the MC samples.

A faster and simpler version of the previous method, called Barlow-Beeston lite, is implemented and commonly used in HistFactory fits. In this version, only one constraint is added per bin, accounting for the total bin content and uncertainty associated to the combination of all the MC samples. On a technical level, the method involves a single-step computation of the relative errors for each bin before the fit, and the computation at every minimisation step of the total constraint parameters per bin, using the previously-obtained relative errors and the updated component fractions. It should be emphasised that, within this method, neither the change in fractions nor the change in component shapes is taken into account in the determination of the relative bin errors.

The RooHammerModel class has been constructed to be used in combination with the Barlow-Beeston lite method, and the way HAMMER is used by the class has been optimised for this approach, as we explain in the following. We define $N_{FF}$ and $N_{WC}$ as the numbers of form-factor parameters and Wilson Coefficients that the user configures to be variable. If both $N_{FF}$ and $N_{WC}$ are relatively large, the number of internal calculations required by HAMMER to process a histogram without evaluating bin errors approximately scales as $(N_{FF} \times N_{WC})^2$. On the other side, if the bin errors also need to be computed, the number of HAMMER calculations approximately scales as $(N_{FF} \times N_{WC})^4$. Given the enormous resource consumption of the second case, we use a strategy to reduce the total number of required computations, exploiting the approximation inherent to the Barlow-Beeston lite method. The strategy consists on doing a double pre-processing of each decay component with HAMMER, producing two separate histograms: a histogram including bin uncertainties but fixed Wilson coefficients and form-factor parameters, and another histogram without bin uncertainties but with the desired parameters set to be variable. The usage of the two histograms together gives all the information needed for fitting, and the required pre-processing time is considerably smaller than that of an alternative single histogram with full features enabled. The histogram without uncertainties will constitute the core of the PDF model, while the histogram with uncertainties will be accessible through a dedicated method of the class called \texttt{getHistogramWithErrors}. As a technical note, the RooHammerModel class assumes that both histograms are present in the same buffer file that results from the HAMMER pre-processing.

The previous setup can be used in a fit configured with HistFactory in the following way: firstly the desired RooHammerModel objects are instantiated and the proxy histogram for each of them is retrieved via  \texttt{getHistogramWithErrors}. These histograms are passed to the HistFactory model as standard templates. Then, the HistFactory model is compiled into a RooFit model, using the dedicated HistFactory methods. This triggers the computation of the relative bin errors used in the Barlow-Beeston lite method, and the creation of the associated nuisance parameters and constraints to be included in the model. Therefore, the relative error parameters  are evaluated at the starting point of the fit.\footnote{This approach assumes the initial values of the parameters not to be too far away from those at the best-fit point.} Finally, the \texttt{RooCustomizer} class allows to substitute the proxy templates in the model with the corresponding RooHammerModel objects, to make use of the shape variation during fitting.

\subsubsection{Uncertainty of the yield parameters.}

If the discussed implementation of the Barlow-Beeston method is used and some of the fit observables are to be interpreted as component yields, an extra consideration should be taken into account, concerning the uncertainty on those parameters.

In HistFactory, any MC-template-based model that describes the histogram of a particular component can be mathematically expressed as:
\begin{equation}
    n(x_b)=\xi\cdot h(x_b),
    \label{eq:histfactory_model_bin}
\end{equation}
where $n(x_b)$ is the expected number of events in a bin of coordinates $x_b$, $\xi$ is a fit parameter and $h(x_b)$ is the MC template, that contains the number of events in the associated MC sample for the considered bin. When performing the fit using the Barlow-Beeston method, the uncertainty on $\xi$ will account for both the statistical fluctuations of the data and the MC sample. It should be noted that $\xi$ does not represent a yield, but a scaling factor to the MC template.

To obtain the associated yield parameter, we can sum Eq.~\ref{eq:histfactory_model_bin} over all the bins, leading to
\begin{equation}
    N_{obs}=\xi\cdot N_{MC},
    \label{eq:histfactory_model_sum}
\end{equation}
where $N_{obs}$ and $N_{MC}$ are, respectively, the total number of events in the data sample and in the MC sample. The measured value for the component yield can be obtained by multiplying $\xi$ and $N_{MC}$. This is typically done in HistFactory by adding multiplicative normalisation factors ($1/N_{MC}$) in the model, that allow to redefine the scaling factor such that its central value corresponds to that of the yield. This is automatically done in RooHammerModel, where the MC-derived PDF is always normalised to one. However, the uncertainty of the yield can not be directly computed as $\sigma(\xi)\cdot N_{MC}$, since $N_{MC}$ has an uncertainty on its own, of statistical nature. Additionally, the uncertainties of $\xi$ and $N_{MC}$ are correlated, since the effect of a statistical fluctuation in $N_{MC}$ is implicitly included in the $\xi$ uncertainty, as part of the fluctuations done by the Barlow-Beeston procedure. As a solution, it is possible to obtain a relation among the uncertainties of the quantities in Eq.~\ref{eq:histfactory_model_sum} by expressing $\sigma(\xi)$ in terms of $\sigma(N_{obs})$ and $\sigma(N_{MC})$, noticing that $N_{obs}$ and $N_{MC}$ are fully uncorrelated:
\begin{equation}
    \sigma(\xi)=\sigma\left(\frac{N_{obs}}{N_{MC}}\right)=\frac{1}{N_{MC}}\sqrt{\sigma(N_{obs})^2+\sigma(N_{MC})^2\frac{N_{obs}^2}{N_{MC}^2}}.
\end{equation}
Re-ordering terms in the previous expression and making use of Eq.~\ref{eq:histfactory_model_sum}, we obtain:
\begin{equation}
    \sigma(N_{obs})=\sigma(\xi)N_{MC}\sqrt{1-\frac{\sigma(N_{MC})^2}{N_{MC}^2}\frac{\xi^2}{\sigma(\xi)^2}}
\end{equation}
The previous expression can then be used to compute the uncertainty on the yield parameter $N_{obs}$ when using the presented implementation of the Barlow-Beeston method, combining the information from the fit parameter $\xi$ and the information from the total number of events in the MC sample (with its uncertainty). Note that the correct uncertainty is always smaller than $\sigma(\xi)N_{MC}$, and the size of the difference between them depends on both the data and MC samples.

\subsection{Passing observable variables to RooHammerModel}

In a typical HistFactory fit, the collection of \texttt{RooRealVar} objects representing the fit variables is automatically constructed when the model is converted into a RooFit one, and can be accessed by the user from that moment onward. These \texttt{RooRealVar} objects have to be passed to the RooHammerModel instances. If this was done in the constructor, as in a typical \texttt{RooAbsPdf}, the RooHammerModel objects would have to be created after the HistFactory model compilation stage. However, in such a scenario the RooHammerModel would not be available at the beginning of the process, to provide the proxy histograms with bin uncertainties (see section~\ref{sec:technical_BB}). With the aim of economising the memory consumption, avoiding duplicated objects, the solution adopted for the RooHammerModel class is to have a constructor (shown in section~\ref{sec:user_RooHammerModel}), where the fit variables are not passed, and then use a devoted method of the class, called \texttt{SetObservables}, to set the variables at the point in the code where they are available. Apart from this default scheme, the RooHammerModel class has an additional constructor, where the fit variables are also passed, in order to operate as usual if used standalone in a generic RooFit fitting framework.

\subsection{Parallelisation}
A generic HAMMER amplitude-reweighting process is divided in two steps: the pre-processing of MC files, not discussed in this document, and the the readout of the histograms, which happens inside the RooHammerModel class.

The first step can be parallelised in terms of MC events, running the HAMMER pre-processing in several CPU processes over separate simulated-data sub-samples, to produce separate HAMMER buffer files. The implementation of the RooHammerModel class allows to profit from this parallelisation scheme, by having a constructor that takes as input a list of buffer-file names (see section~\ref{sec:user_RooHammerModel}). The information from the different files will be internally combined by summing the histograms, using the corresponding HAMMER functionality~\cite{Bernlochner:2020tfi}.

Concerning the parallelisation at histogram-readout time, the most relevant possibility is the parallelisation of the likelihood evaluation over different sets of bins. This can be done using the standard method \texttt{RooFit::NumCPU} in the fit, that allows the users to specify the number of processes to be run in parallel. It should be noted that the parallelisation done with this method relies on multi-processing, not on multi-threading.\footnote{Note that the HAMMER tool provides a set of thread-safe specific methods that are useful for multi-threading parallelisation schemes.} This implies that the parallel workers load internally independent instances of the likelihood objects, including the HAMMER objects. On one side, this guarantees code safety under parallelisation, since there is no cross-talk between the different processes. On the other side, the multiple loading of the objects implies a higher memory consumption for a parallel fit, which should be considered by the user.
\section{Testing the framework}
\label{sec:tests}

This section presents a set of generate-and-fit pseudo-experiments, called ``toy'' experiments, aimed at checking the fitting procedure using the new RooHammerModel class and the techniques explained in the previous section in a typical fit configured with HistFactory. The experiments are designed to test a realistic fitting situation from a technical viewpoint, with a simple physics scenario.

\subsection{Technical setup of the experiments}

For each pseudo-experiment, the event generation of all the needed samples is done using RapidSim \cite{Cowan_2017}, with a LHCb configuration, and EvtGen \cite{Lange:2001uf}, although the conclusions of the checks are general to any experimental setup. The samples whose decay amplitude is to be re-weighted are pre-processed with HAMMER, to produce the corresponding HAMMER buffer files. A fit is done using a set of the samples (see the next sub-section) to construct the physics dataset and another set to construct the fitting model, making use of RooHammerModel when relevant. The previously discussed implementation of the Barlow-Beeston lite method is used, to account for the uncertainty on the templates.

The fit results are collected for a large number of pseudo-experiments, and their statistical agreement with the generated parameter values and uncertainties is used to check the validity of the framework.

\subsection{Benchmark physics scenario}

As a particular physics scenario, a combination of two decay channels is chosen:\footnote{Charge conjugation is implied in what follows.} $\overline{B}^0\to D^{*+}\mu^-\overline{\nu}_{\mu}$ and $\overline{B}^0\to D^{*+}\tau^-\overline{\nu}_{\tau}$, with $D^{*+}\to D^+\pi^0$ and $\tau^-\to\mu^-\overline{\nu}_{\mu}\nu_{\tau}$. It is assumed for this scenario that neither the neutrinos nor the $\pi^0$ are reconstructed experimentally, so the visible final state, common to both the muonic and tauonic decays, is $D^+\mu^-$.

To investigate this combination of decays, three variables are used in the fit: the muon energy in the $\overline{B}^0$ rest frame, $q^2=(p(\overline{B}^0)-p(D^+))^2$ and $M_{miss}^2=(p(\overline{B}^0)-p(D^+)-p(\mu^-))^2$, where $p(P)$ denotes the four-momentum of particle $P$. No experimental resolution effects are considered in these studies, although the usage of the $D^+$ to compute the previous variables instead of the $D^{*+}$ leads to a broadening of the distributions, because of the non-reconstructed $\pi^0$.

For each experiment, four samples with the previous decays are generated: \texttt{muonic\_data} and \texttt{tauonic\_data}, that are combined to represent the physics dataset, and \texttt{muonic\_model} and \texttt{tauonic\_model}, that provide the templates for the fit model. The \texttt{muonic\_data} sample is generated using the CLN parameterisation~\cite{Caprini_1998}, while the \texttt{muonic\_model} sample is generated with a pure phase-space model. Both the \texttt{tauonic\_data} and \texttt{tauonic\_model} samples are generated with a pure phase-space model. In the experiments, HAMMER is used to reweight \texttt{muonic\_model} from phase space to CLN. Apart from generating different events for each experiment, the total number of events in each sample, $N_{muonic}^{data}$, $N_{tauonic}^{data}$, $N_{muonic}^{model}$ and $N_{tauonic}^{model}$, is also fluctuated, according to a poissonian distribution.

The parameters to be measured in each fit are: the combined yield, $N_{total}=N_{muonic}^{data}+N_{tauonic}^{data}$; the ratio of yields, $R(D^*)=N_{tauonic}^{data}/N_{muonic}^{data}$, and three form-factor parameters of the CLN parameterisation, $\rho^2$, $R_1$ and $R_2$. Since the CLN form factors are not linear on the corresponding parameters~\cite{Caprini_1998}, the presented toy study also tests the validity of the linear approximations done by HAMMER (see section~\ref{sec:user}). As a technical note, since the form-factor parameters in HAMMER are constructed as differences with respect to central values set by the user at the pre-processing step, the three parameters that are actually measured in the fit correspond to those differences, denoted by $\Delta\rho^2$, $\Delta R_1$ and $\Delta R_2$.

\textbf{Generation values.} In each experiment, the muonic and tauonic samples constituting the physics dataset are produced using the generation values $N_{total}=10^6$ events, $R(D^*)=0.3$ and $\Delta\rho^2=\Delta R_1=\Delta R_2=0$, with central form-factor parameter values $\rho^2=1.207$, $R_1=1.401$ and $R_2=0.854$. The respective yields of the muonic and tauonic samples used to construct the fitting model are generated with the same central values as their physics-dataset counter-parts, such that the parameter uncertainties originating from both the physics dataset and the model templates are comparable in size.

\subsection{Results of the validation}

A set of 1000 toy experiments have been performed with the previous setup, leading to 995 successful fits. In each fit, the pull variable for each parameter $\theta$ is evaluated as follows:
\begin{equation}
    \theta_{pull} = \frac{\hat{\theta} - \theta_{true}}{\sigma_{\theta}},
\end{equation}
where $\theta_{true}$ is the value used for generation, and $\hat{\theta}$ and $\sigma_{\theta}$ are the fit estimates for the value and uncertainty, respectively. An exception is done for the yield-like parameters $N_{total}$ and $R(D^*)$, for which $\sigma_{\theta}$ is computed as explained at the end of section~\ref{sec:technical_BB}.

\begin{figure}[t]
    \centering
    \includegraphics[width=0.48\textwidth]{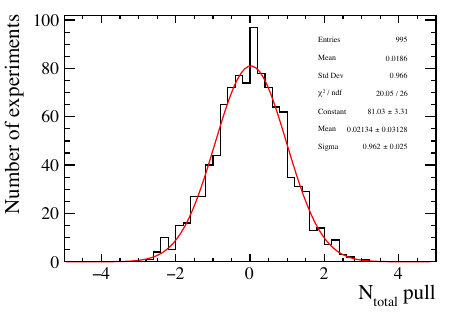}
    \includegraphics[width=0.48\textwidth]{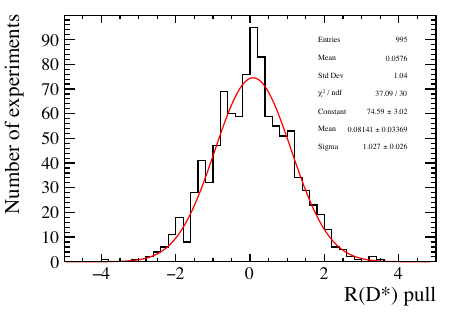}\\
    \includegraphics[width=0.48\textwidth]{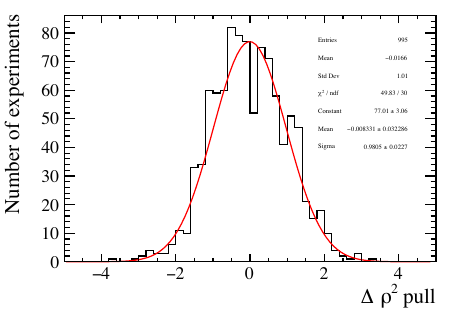}
    \includegraphics[width=0.48\textwidth]{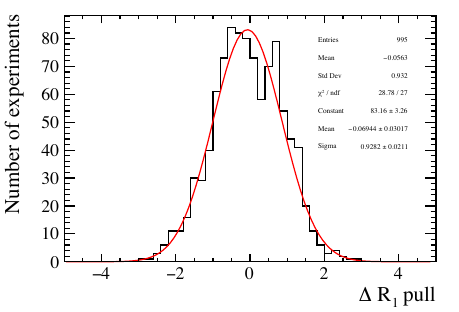}\\
    \includegraphics[width=0.48\textwidth]{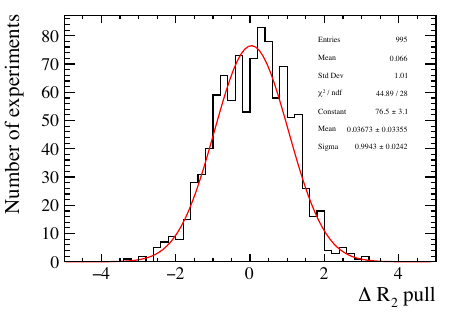}
     \caption{Pull distributions for the fit parameters, corresponding to the ensemble of toy experiments used to validate the framework. The result of a gaussian fit to each distribution is superimposed as a red curve.\vspace{4pt}}
    \label{fig:tests_pulls}
\end{figure}

The obtained pull distributions for the successful fits are shown in figure~\ref{fig:tests_pulls}. For all the distributions, the parameters of a Gaussian fit are consistent with those of a standard Gaussian within three standard deviations. This demonstrates that, in the studied configuration, the fit provides unbiased estimates for the parameters, with the correct coverage properties.

\section{Performance assessment}
\label{sec:performance}

The average CPU time per minimization step is used to asses the performance.
For this study, the chosen benchmark model consists on a single component, $\overline{B}^0\to D^{*+}\mu^-\overline{\nu}_{\mu}$, which is represented by a RooHammerModel object. To focus on a computationally more involved case from the HAMMER side, instead of the CLN parameterisation previously inspected we now use the BGL222 scheme presented in Ref.~\cite{Belle:2017rcc}, allowing us to float six form factor parameters in the fit. The Wilson Coefficients are fixed to their SM values, and the total yield is also fixed. Having form factor parameters as the only fit variables ensures that the internal reweighting performed by the RooHammerModel class occurs at every step, since caching of the histogram shape is not possible in that configuration. The number of bins used in the model is 3600. The minimisation of the likelihood function is performed by Minuit~\cite{James:1994vla}, which is configured through the \texttt{RooMinuit} class in RooFit. The study is done on a CentOS-7-x86\_64 architecture, using a 2.2 GHz Intel Core Processor (Broadwell, IBRS). Regarding software versions, RooHammerModel v1.1 is used, in combination with HAMMER v1.1.0 and ROOT v6.16.

In the first place, we evaluate the average time per minimisation step taken by the fit when a single process is used and the Barlow-Beeston method is deactivated, finding a value of 48 ms. If the likelihood evaluation is parallelised over four processes, the time is reduced by a modest 20\%. This result points to the internal reading of the HAMMER histogram by RooHammerModel having a dominating effect, since this action does not profit from the parallelisation, while the per-bin evaluation of the already read histogram is much less time consuming. It should be noted that, in a more general model with floating parameters other than the HAMMER related ones, the automatic histogram-caching functionality of RooHammerModel will reduce the relative importance of the histogram-reading from HAMMER, reducing the average time per minimisation step and improving its scaling with parallelisation. Finally, we evaluate the average time per minimisation step when the Barlow-Beeston method is activated. In that configuration, a time of 130 ms is found when using a single process, and this metric is not found to improve with parallelisation.
\section{Conclusions}
\label{sec:conclusions}
This article describes an interface of the HAMMER tool with the well known data-analysis framework HistFactory. The interface, based on C++, is generic for any decay channel, has a user-friendly interaction and can be integrated in plain RooFit if desired. It has been optimised in terms of speed and memory handling, and has been tested in detail. The particular solutions adopted for the data analysis and computing challenges encountered when developing the interface are carefully described. We hope that they might serve as an inspiration to other groups trying to interface HAMMER with other fitting frameworks. The code of the interface is provided in the following open-access Gitlab repository: \href{https://gitlab.cern.ch/InterfacingHammer/roohammermodel}{https://gitlab.cern.ch/InterfacingHammer/roohammermodel}. The description and studies presented in this paper correspond to RooHammerModel v1.1.

\acknowledgments

The RooHammerModel interface has been developed in continuous symbiosis and interaction with the developers of the HAMMER tool. We would especially like to thank Michele Papucci and Dean Robinson, for all the constructive discussions, the detailed guidance on the usage of their tool and for the implementation of extra features in HAMMER that allowed this interface to work as needed. We would also like to thank Phoebe Hamilton, for the useful advice on technical aspects of the HistFactory framework. Finally, we want to express our gratitude to the other members of the Semileptonic Working Group of the LHCb Collaboration, for the discussions that helped to improve this interface.


\end{document}